\documentclass[twocolumn,showpacs,preprintnumbers,amsmath,amssymb,superscriptaddress]{revtex4}

\usepackage{graphicx}
\usepackage{dcolumn}
\usepackage{bm}

\def\lsi{\raise0.3ex\hbox{$<$\kern-0.75em\raise-1.1ex\hbox{$\sim$}}}
\def\gsi{\raise0.3ex\hbox{$>$\kern-0.75em\raise-1.1ex\hbox{$\sim$}}}
\newcommand{\lsim}{\mathop{\lsi}}

\newcommand{\src}{\mbox{\footnotesize{src}}}
\newcommand{\obs}{\mbox{\footnotesize{obs}}}

\begin{document}

\title{Gravity-induced birefringence within the framework
         of Poincar\'e gauge theory}
\author{Oliver Preuss}
\affiliation{Max-Planck-Institut f\"ur Sonnensystemforschung,
             D-37191 Katlenburg-Lindau, Germany}
\author{Sami K. Solanki}
\affiliation{Max-Planck-Institut f\"ur Sonnensystemforschung,
         D-37191 Katlenburg-Lindau, Germany}
\author{Mark P. Haugan}
\affiliation{Department of Physics,
             Purdue University 1396, West Lafayette, Indiana 47907, USA}
\author{Stefan Jordan}
\affiliation{Astronomisches Rechen-Institut, ZAH, D-69120 Heidelberg, Germany}    
\date{\today}

\begin{abstract}
    Gauge theories of gravity provide an elegant and promising
    extension of general relativity. In this paper we show that the
    Poincar\'e gauge theory exhibits gravity-induced birefringence under
    the assumption of a specific gauge invariant nonminimal coupling between
    torsion and Maxwell's field. Furthermore we give for the first
    time an explicit expression for the induced phaseshift between two
    orthogonal polarization modes within the Poincar\'e framework. Since
    such a phaseshift can lead to a depolarization of light emitted from
    an extended source this effect is, in principle, observable. We use
    white dwarf polarimetric data to constrain the essential coupling
    constant responsible for this effect.
\end{abstract}
\maketitle
\section{Introduction}
    Almost 90 years after its formulation, Einstein's concept of gravity
    as a purely geometrical property of a four dimensional Riemannian manifold
    still provides a valid description of gravitational interactions. A major
    reason for this success is that the influence of matter is introduced solely
    by means of its energy-momentum tensor. It is clear that this phenomenological
    approach is justified as long as we are interested only in macroscopic
    events but obviously a more complete description of matter properties is achieved
    if we include also spin angular momentum besides energy momentum as an additional
    basic feature which determines the dynamics of matter on microscopic scales.

    Currently, in this sense the most promising extensions of general relativity
    are given in the framework of gauge theories of gravity \cite{gh96,bl02}.
    The description of fundamental interactions by means of gauge symmetries has
    become a cornerstone in modern theoretical physics. Especially Poincar\'e
    symmetry has been proven to play an important role in particle physics and
    the results of the Colella-Overhauser-Werner (COW) experiment \cite{cow} lead
    to consider Poincar\'e gauge theory (PGT) with a Riemann-Cartan spacetime $U_4$
    as a very natural alternative to general relativity \cite{h97}. PGT features
    torsion and curvature as gravitational gauge fields so that within this
    framework both mass and spin act as sources of the gravitational field.

    In this paper we focus on consequences which arise from a possible nonminimal
    coupling between the torsion of PGT and the electromagnetic field. In contrast
    to the usual minimal coupling scheme where the propagation of electromagnetic waves
    is not affected by the presence of torsion, the direct coupling of the electromagnetic
    field with a gravitational gauge field leads to new physical effects like
    gravity-induced birefringence \cite{sh96,s99,sp03,p03,p04}. This nonminimal approach
    is motivated also by low-energy limits of string theories where torsion is identified
    with a massless antisymmetric second rank Kalb-Ramond (KR) field, present in most
    supergravity theories and as such in the massless sector of the most viable string
    theories \cite{green}. Consequently, the covariant derivative of this KR-field is
    a field of the same tensor type as the torsion field we consider and so can, in
    principle, couple to the electromagnetic field in the same ways that torsion can \cite{kar}.
    Recently, Laemmerzahl \& Hehl investigated light propagation within a Finslerian
    Geometry of spacetime \cite{lh04}. They found that vanishing birefringence
    automatically yields a Riemannian structure and no Finslerian structure can occur.
    
    In addition to the conventional Maxwell Lagrangian, the specific nonminimal coupling
    we employ is given by
    \begin{equation}\label{l1}
     L_{EM}=p^2\star(T_{\alpha}\wedge F)T^{\alpha}\wedge F\quad,
    \end{equation}    
    where $p$ denotes a coupling constant with the dimension of length, $\star$ is the
    Hodge dual, $T$ denotes the torsion and $F$ the electromagnetic field which is related
    in the usual way to its potential $A$ \cite{sp03,p03,p04,p04a}. This addition is gauge
    invariant and, so, compatible with charge conservation. In this context, it was later shown
    by Itin \& Hehl \cite{itin} that (\ref{l1}) is a special case of a complete family of
    quadratic torsion lagrangians which couple to Maxwell's field and which leads besides
    birefringence to an axion-induced optical activity of spacetime and, furthermore, to
    a torsion dependent speed of light:
    \begin{equation}
      L_{EM}=-\frac{1}{8}p\sum_{a,\ldots,q}(F_{ab}F_{cd}T_{klm}T_{npq})\quad .
    \end{equation}
    Here the summation is performed by contracting the indices by means of the metric tensor.  
    
    This paper is organized as follows: First we give a brief recapitulation of the Poincar\'e
    gauge theory where the field strengths of the compensating gauge fields are identified as
    torsion and curvature. Then, using the Baekler-Lee solution for a spherically symmetric
    torsion we show that the Lagrangian (\ref{l1}) leads to gravity-induced birefringence
    and give an explicit expression for the accumulated phase shift between orthogonal
    polarization modes in the gravitational field of a star. Since this effect leads to a
    depolarization of radiation, emitted from an extended source we use white dwarf polarimetric
    data to set strong limits on the essential coupling constant. Finally, a discussion and
    conclusions are presented.
  \section{Poincar\'e Gauge Theory}
    Our subsequent brief summary of basic features of PGT mainly follows the notation
    from Blagojevi\'c \cite{bl02}. PGT belongs to the class of Lagrangian-based
    theories of gravity which means that the equations of motion are given by the
    Euler-Lagrange equations of the action integral
    \begin{equation}\label{action}
      S_M = \int\,d^4x\,{\cal L}_M(\phi,\partial\phi)\quad ,
    \end{equation}
    with the matter field $\phi(x)$ as the dynamical variable. To ensure the conservation
    of energy momentum and angular momentum one demands the invariance of (\ref{action})
    under global Poincar\'e transformations
    \begin{equation}
      x'^{\mu}=x^{\mu}+\xi^{\mu}(x), \quad \xi^{\mu}=\omega^{\mu}{}_{\nu}x^{\nu}+\epsilon^{\mu}
    \end{equation}
    where the Lorentz rotations $\omega^{\mu\nu}=-\omega^{\nu\mu}$ and translations
    $\epsilon^{\mu}$ provide ten constant parameters. Here, greek indices alway refer to
    coordinate lines of the underlying Minkowski space $M_4$. As a consequence, matter
    fields $\phi(x)$ transform according to
    \begin{equation}
      x'^i=x^i+\omega^i{}_jx^j+\epsilon^i ,\quad
      \phi'(x')=(1+\frac{1}{2}\omega^{ij}\Theta_{ij})\phi(x)\quad ,
    \end{equation}
    where $\Theta_{ij}$ denotes the spin matrix, related to the multicomponent structure
    of $\phi(x)$. Latin indices refer to a local Lorentz frame, tangent to $M_4$. If one defines
    $\delta_0\phi(x)=\phi'(x)-\phi(x)$, the action (\ref{action}) is invariant under the
    transformation $x'+x+\xi(x)$ if
    \begin{equation}\label{trans}
      \Delta{\cal L}\equiv\delta_0{\cal L}+\xi^{\mu}\partial_{\mu}{\cal L}+(\partial_{\mu}
      \xi^{\mu}){\cal L}=0\quad ,
    \end{equation}
    where $\delta_0{\cal L}=(\partial{\cal L}/\partial\phi)\delta_0\phi+(\partial{\cal L}/
    \partial\phi_{,k})\delta_0\phi_{,k}$ \cite{ki61,ut56}, so that Noether's theorem leads to
    conserved energy momentum and angular momentum tensors.

    In a next step, the Poincar\'e transformations are generalized by replacing the ten
    constant group parameters with corresponding functions of spacetime points, i.e.
    \begin{equation}
     \omega^{ij}\rightarrow\omega^{ij}(x),\quad \epsilon^k\rightarrow\epsilon^k(x)\quad.
    \end{equation}
    Based on experience, e.g. from QED, it is then no surprise that the invariance
    condition (\ref{trans}) is now violated. However, this problem can be circumvented in the
    usual way by means of a covariant derivative $\nabla_k\phi$ of $\phi$ which is introduced
    in two steps
    \begin{eqnarray}
      \nabla_{\mu}\phi&=&(\partial_{\mu}+A_{\mu})\phi,\quad A_{\mu}\equiv\frac{1}{2}A^{ij}{}_{\mu}
      \Theta_{ij}\quad ,\\
      \nabla_k\phi&=&\delta^{\mu}_k\nabla_{\mu}\phi-A^{\mu}{}_k\nabla_{\mu}\phi\equiv h_k{}^{\mu}
      \nabla_{\mu}\phi\quad ,
    \end{eqnarray}
    with the new field $h_k{}^{\mu}=\delta^{\mu}_k-A^{\mu}{}_k$. In order to restore the local
    invariance of the theory, one introduces $\tilde{{\cal L}}_M=\Lambda{\cal L}_M(\phi,\nabla_k\phi)$,
    where $\Lambda$ is a suitable function of the new fields. Then the invariance condition
    (\ref{trans}) is restored if $\delta_0\Lambda+\partial_{\mu}(\xi^{\mu}{\Lambda})=0$, which is
    given by $\Lambda=\det(b^k{}_{\mu})\equiv b$, where $b^k{}_{\mu}$ is just the inverse of
    $h_k{}^{\mu}$: $b^k{}_{\mu}h_i{}^{\mu}=\delta^k_i$.
    Finally, the locally invariant Lagrangian for matter fields reads
    \begin{equation}
      \tilde{{\cal L}}_M=b{\cal L}_M(\phi,\nabla_k\phi)\quad .
    \end{equation}
    The corresponding field strengths of the new compensating fields $b^k{}_{\mu}$ and $A^{ij}{}_{\mu}$
    are given by the tensors
    \begin{eqnarray}
      F^{ij}{}_{\mu\nu}&\equiv& \partial_{\mu}A^{ij}{}_{\nu}-\partial_{\nu}A^{ij}{}_{\mu}\\\nonumber
      & &+A^i{}_{s\mu}A^{sj}{}_{\nu}-A^i{}_{s\nu}A^{sj}{}_{\mu} \\
      F^i{}_{\mu\nu}&\equiv&\nabla_{\mu}b^i{}_{\nu}-\nabla_{\nu}b^i{}_{\mu} \quad ,
    \end{eqnarray}
    which are called the Lorentz and translation field strengths, respectively. From the structure
    of these tensors it is now easy to conclude that the translation field strength $F^i{}_{\mu\nu}$
    is nothing but the torsion $T^{\lambda}{}_{\mu\nu}$, while the Lorentz field strength
    $F^{ij}{}_{\mu\nu}$ can be identified with the curvature $R^{\lambda}_{\tau\mu\nu}$ \cite{bl02}. Therefore,
    it is evident that PGT possesses a Riemann-Cartan spacetime where both mass {\em and} spin are
    sources of the gravitational field.   
  \section{Birefringence Analysis}
    The simplest solution of the PGT field equations with dynamic torsion is given by the
    spherically symmetric Baekler-Lee solution \cite{bak81,bl83}.

    Starting with the usual Schwarzschild tetrad
    \begin{eqnarray}
      e^{\hat{t}}&=&\sqrt{2\Phi}\,dt,\quad e^{\hat{r}}=dr/\sqrt{2\Phi} \\ e^{\hat{\theta}}&=&r\, d\theta,
      \quad\quad\,\; e^{\hat{\phi}}=r \sin\theta\,d\phi\quad ,
    \end{eqnarray}
    the Baekler-Lee solution is appreciably simplified by applying a suitable boost
    $\vartheta^{\alpha}=\Lambda^{\alpha}_{\,\beta}\,e^{\beta}$ so that the corresponding
    orthogonal coframe takes the form \cite{gh96}
    \begin{eqnarray}\label{cf}
      \vartheta^{\hat{t}} &=& \frac{1}{2}((\Phi+1)dt+(1-\frac{1}{\Phi})dr) \\
      \vartheta^{\hat{r}} &=& \frac{1}{2}((\Phi-1)dt+(1+\frac{1}{\Phi})dr) \nonumber\\
      \vartheta^{\hat{\theta}} &=& r\, d\theta \nonumber\\
      \vartheta^{\hat{\phi}} &=& r\sin\theta\, d\phi\nonumber\quad ,
    \end{eqnarray}
    with
    \begin{equation}
      \Phi:=1-\frac{2(Mr-q^2)}{r^2}-\frac{\kappa}{4\ell^2}r^2 \quad.
    \end{equation}
    Here $M$ and $q$ denote the gravitational mass and charge, respectively. $\ell =$ Plancklength and
    $\kappa$ is an additional coupling constant \cite{gh96}.
    The corresponding metric is then given by
    \begin{equation}
      ds^2 = -\Phi\, dt^2+\frac{1}{\Phi}\,dr^2+r^2(d\theta^2+\sin^2\theta\,d\phi^2) \quad .
    \end{equation}
    Now, the torsion of the Baekler-Lee solution reads 
    \begin{eqnarray}\label{pgttorsion}
     T^{\hat{t}}=T^{\hat{r}} &=& \frac{Mr-2q^2}{r^3}\, \vartheta^{\hat{t}}\wedge\vartheta^{\hat{r}}\,,\nonumber \\
     T^{\hat{\theta}} &=& \frac{Mr-q^2}{r^3}\left(\vartheta^{\hat{t}}\wedge\vartheta^{\hat{\theta}}
                          -\vartheta^{\hat{r}}\wedge\vartheta^{\hat{\theta}}\right)\,, \\
     T^{\hat{\phi}} &=& \frac{Mr-q^2}{r^3}\left(\vartheta^{\hat{t}}\wedge\vartheta^{\hat{\phi}}
                        -\vartheta^{\hat{r}}\wedge\vartheta^{\hat{\phi}}\right)\quad .\nonumber			
    \end{eqnarray}
    This solution is consistent with the most general static, spherically $O(3)$-symmetric form for a
    torsion field \cite{tr95}
    \begin{eqnarray}
      T^0 &=& \alpha(r)\,\theta^0\wedge\theta^1
              +\tilde{\alpha}(r)\,\theta^2\wedge\theta^3\quad , \\
      T^1 &=& \beta(r)\,\theta^0\wedge\theta^1 
              +\tilde{\beta}(r)\,\theta^2\wedge\theta^3\quad , \\
      T^2 &=& \gamma_{(1)}\,\theta^0\wedge\theta^2
              +\gamma_{(2)}\,\theta^0\wedge\theta^3 \nonumber\\
              &&{}+\gamma_{(3)}\,\theta^1\wedge\theta^2 
              +\gamma_{(4)}\,\theta^1\wedge\theta^3 \quad , \\
      T^3 &=& \gamma_{(1)}\,\theta^0\wedge\theta^3
              -\gamma_{(2)}\,\theta^0\wedge\theta^2 \nonumber\\
              &&{}+\gamma_{(3)}\,\theta^1\wedge\theta^3 
              -\gamma_{(4)}\,\theta^1\wedge\theta^2 \quad .
    \end{eqnarray}
    The solution (\ref{pgttorsion}) is a special case having 
    $\tilde{\alpha}(r)=\tilde{\beta}(r)=\gamma_{(2)}=\gamma_{(4)}=0$. Plugging this general torsion
    field into the Lagrangian density (\ref{l1}) the coefficients of the magnetic and electric field
    components can be expressed in terms of SO(3)-symmetric tensors $\xi^{ij},\,\zeta^{ij}$ and
    $\gamma^{ij}$ which represent spatial anisotropy induced by the gravitational field. As shown in
    \cite{sp03,p03} the accumulated phaseshift $\Delta\Phi$ which is due to the fractional difference
    $\delta c/c$ in the propagation speed of linear polarization states with frequency $\omega$
    \begin{equation}
      \Delta\Phi = \omega\int\frac{\delta c}{c}\,dt \quad ,
    \end{equation}
    can be expressed in terms of the spherical components of these SO(3) tensors by using the
    Haugan-Kauffmann formalism \cite{hk95}. However, one has to be careful since \cite{hk95}
    uses a $(+---)$ metric, while the Baekler-Lee solution is based on a $(-+++)$ metric
    so that either (\ref{pgttorsion}) or the Haugan-Kauffmann formalism has to be rewritten in
    terms of a different metric.
  
    The general expression for $\delta c/c$ is then given by
    \begin{equation}\label{deltac2}
      \frac{\delta c}{c} = \sqrt{\frac{2}{3}}\,\sin^2\theta\,\sqrt{\left(
      \xi_0^{(2)}+\zeta_0^{(2)}\right)^2+4\left(\gamma^{(2)}_0\right)^2} \quad ,
    \end{equation}    
    with
    \begin{eqnarray}
      \xi^{(2)}_0 &=& (\gamma^2_{(3)}+\gamma^2_{(4)}) \\
      \zeta^{(2)}_0 &=& -(\alpha^2-\beta^2)+2(\gamma^2_{(1)}+\gamma^2_{(2)})\\
      \gamma^{(2)}_0 &=& (\gamma_{(1)}\gamma_{(4)}-\gamma_{(2)}\gamma_{(3)}) \quad .
    \end{eqnarray}
    Comparing the coefficients of the general $O(3)$-symmetric Torsion with the Baekler-Lee solution we find  
    \begin{eqnarray}
      \xi^{(2)}_0 &=& \gamma^2_{(3)} = \frac{\left(Mr-q^2\right)^2}{r^6} \\
      \zeta^{(2)}_0 &=& \gamma^2_{(1)} =  \frac{\left(Mr-q^2\right)^2}{r^6} \\
      \gamma^{(2)}_0 &=& 0\quad ,
    \end{eqnarray}
    which leads to
    \begin{eqnarray}
      \frac{\delta c}{c} &=& \sqrt{\frac{2}{3}}\,\sin^2\theta\,\sqrt{\left(2\frac{\left(Mr-q^2\right)^2}{r^6}
      \right)^2} \\
      &=& 2\sqrt{\frac{2}{3}}M^2\sin^2\theta\frac{1}{r^4}\quad ,
    \end{eqnarray}
    in case of vanishing charge $q = 0$. Therefore, the total phase shift becomes
    \begin{equation}\label{dp1}
      \Delta\Phi=2\omega\sqrt{\frac{2}{3}}M^2\int_{t_0}^{t_1}\frac{\sin^2\theta}{r^4}\,dt\quad .
    \end{equation}
    The evaluation of this integral requires a ray parametrization ${\bf x}(t)={\bf b}+{\bf k_0}t$
    where the unit vector ${\bf k_0}$ denotes the ray direction and ${\bf b}$ is the impact vector
    that connects the center of the star with the closest point on the ray. When ${\bf b}$ is
    smaller than the radius $R$ of the star, the portion of the ray inside the object is, of course,
    of no interest. The integration of (\ref{dp1}) is performed from the star's surface with
    $t_0=(R^2-b^2)^{1/2}$ along a straight line up to an observer at an infinite distance $t_1=\infty$,
    which yields
    \begin{equation}
      \Delta\Phi=2\sqrt{\frac{2}{3}}\omega M^2 R^2(1-\mu^2)\int_{t_0=R\mu}^{\infty}\frac{dt}{(R^2(1-\mu^2)+t^2)^3}\quad ,
    \end{equation}
    where $\mu$ denotes the cosine of the heliocentric angle $\theta$ between the ray's source and the center
    of the visible stellar disc. This integral is easily evaluated, so that we finally get the total phase shift which
    accumulates between two orthogonal polarization states within the framework of Poincar\'e gauge theory
    \begin{eqnarray}\label{dphi}  
      \Delta\Phi &=& \sqrt{\frac{2}{3}}\frac{4\pi M^2 p^2}{\lambda R^3}\left(\frac{3\pi}{16(1-\mu^2)^{3/2}}\right.\\\nonumber
                  &-&\left.\frac{\mu}{4}-\frac{3\mu}{8(1-\mu^2)}
                 -\frac{3}{8(1-\mu^2)^{3/2}}\arcsin(\mu)\right)   
    \end{eqnarray}
    with the new Poincar\'e coupling constant $p$ having the dimension of length and
    the mass $M$ in geometrized units. It is interesting to see that (\ref{dphi}) shows
    a remarkable similarity to the phase shift formula from Moffat's old version of NGT \cite{g91}.
    Nevertheless, while this nonvanishing birefringence was found on the basis of the special
    Baeckler-Lee solution, it is important to note that Rubilar et al. \cite{r03} proved that
    our nonminimal Lagrangian (\ref{l1}) leads to birefringence even for a general
    O(3)-symmetric torsion field. 
  \section{Astrophysical constraints}
    The question how gravitational birefringence influences polarized radiation
    in the vicinity of a particular star depends, among others, on the properties
    of the emitting source. In case of a pointlike source of polarized radiation, 
    all received light suffers the same phase shift $\Delta \Phi(\mu)$. 

    Polarized light is described by means of wavelength dependend Stokes parameters
    $I_{\lambda},\,Q_{\lambda},\,U_{\lambda},V_{\lambda}$ \cite{s62}, where Stokes $Q$ represents
    the difference between linear polarization parallel and perpendicular to the local
    stellar limb. The effect of gravitational birefringence is to introduce a crosstalk
    between the linear polarization parameter Stokes $U$ and the net circular polarization,
    $V$. This crosstalk is such that although the observed values $U_{\obs}$ and $V_{\obs}$
    differ from the values emitted by a point source, $U_{\src}$ and $V_{\src}$, the
    composite degree of polarization remains equal: $(U_{\obs}^2 + V_{\obs}^2)^{1/2} =
    (U_{\src}^2 + V_{\src}^2)^{1/2}$.

    In case of an extended source covering a range of $\mu$ values, light emitted from
    different points suffers different phase shifts and, so, adds up to an incoherent
    superposition. Using the additive properties of Stokes parameters, summing over
    different contributions yields a reduction of the observed polarization relative
    to the light emitted from the source:
    $(U_{\obs}^2 + V_{\obs}^2)^{1/2} < (U_{\src}^2 + V_{\src}^2)^{1/2}$. 
    Since the rotationally modulated polarization from magnetic white 
    dwarfs can only be produced by an extended source \cite{chile}, any 
    observed (i.e. non-zero) degree of polarization provides a limit 
    on the strength of birefringence induced by the star's gravitational 
    field \cite{s99}. 

    It is generally agreed that polarized radiation from white dwarfs
    is produced at the stellar surface as a result of the presence of ultrastrong
    (up to $10^5$ T) magnetic fields \cite{lan92}. Since the disk of a white dwarf is
    unresolved, only the total polarization from all surface elements is observable.
    Therefore, the flux of net circular polarization at wavelength $\lambda$ emitted
    toward the observer can be written as
    \begin{equation}\label{vl}
      V_{\lambda, {\rm tot}}(p) = 2\pi\int\int \,V_{\lambda}(\mu,B,\theta)\,\cos(\Delta\Phi)
      \, \mu \, d\theta\, d\phi \,\, \quad . 
    \end{equation}         
   
    \noindent Here, the Stokes parameter $V_{\lambda}$ changes over the
    visible hemisphere and depends on the wavelength $\lambda$, the location $\mu$ 
    (limb darkening), the total magnetic field strength $B$, the angle $\theta$ between
    the magnetic field and the line-of-sight component, and on the parameters of the 
    stellar atmosphere influencing line formation. The influence of gravitational birefringence
    on the polarization is introduced by the term $\cos(\Delta\Phi)$ as a function of $\mu$.
    The Stokes parameters can be calculated by solving the radiative transfer equations
    through a magnetized stellar atmosphere on a large number of surface elements on the
    visible hemisphere (e.g. \cite{j92}). If the star is rotating, the spectrum and polarization 
    pattern changes according to the respective magnetic field distribution 
    visible at a particular moment. The degree of circular polarization, is obtained by
    dividing Eq.~(\ref{vl}) by the total stellar flux $I_{\lambda, {\rm tot}}$ emitted to
    the observer at wavelength $\lambda$. 
    Below we will calculate a maximum circular polarization $V_{\lambda, {\rm max}}/
    I_{\lambda, {\rm tot}}$ from radiative transfer calculations which is  
    higher than the observed value $V_{\lambda, {\rm obs}}/I_{\lambda, {\rm obs}}$. 
    Then we assume that the reduction from  $V_{\lambda, {\rm max}}$ to  $V_{\lambda, 
    {\rm obs}}$ is entirely due to the factor $\cos(\Delta\Phi(p))$ in Eq.~(\ref{vl}), 
    thereby calculating the maximum value for $p$, i.e. our limit on $p$ is reached 
    as soon as $V_{\lambda, {\rm tot}}/I_{\lambda, {\rm tot}}$ in Eq.~(\ref{vl}) becomes 
    smaller than $V_{\lambda,{\rm obs}}/I_{\lambda, {\rm tot}}$ for a certain value 
    of $p$. 
   
    $\mbox{RE J0317-853}$ is a highly unsual object within the class of isolated 
    magnetic white dwarfs which sets several records: Besides being the most rapidly rotating star 
    ($P=725$ sec) of this type it is also the most massive at $1.35\,M_{\odot}$, 
    close to the Chandrasekhar limit \cite{b95} with a corresponding radius of 
    only $0.0035\, R_{\odot}$. In \cite{jb99} a degree $V_{\lambda, {\rm obs}}/I_{\lambda, 
    {\rm tot}}$ of  $20$\%  at $\lambda= 576$ nm \cite{jb99}, $\mbox{RE J0317-853}$ 
    is also the magnetic 
    white dwarf with the highest known level of circular polarization. Due 
    to its small radius and high degree of circular polarization, $\mbox{RE J0317-853}$ is a 
    very suitable object for setting limits on gravitational birefringence.
    The analysis of time resolved UV flux spectra obtained with the Hubble Space 
    telescope has shown that the distribution of the field moduli
    is approximately  that of an off-centered magnetic dipole oriented obliquely to 
    the rotation axis with a polar field strength at the surface of $B_d=3.63\cdot10^4$ T, 
    leading to visible surface field strengths between $1.4\cdot 10^4$ T and 
    $7.3\cdot 10^4$ T \cite{bjo99}. This model is not only able to describe the UV, 
    but also  the optical spectra (Jordan et al. in prep.), which means that the 
    distribution of the magnetic field moduli - but not necessaryly of the longitudinal 
    components -  is correctly described. This result is completely independent of the 
    magnitude of the gravitational birefringence, since it is obtained entirely from the
    intensity spectrum. 
    From radiative transfer calculations it follows that at the phase of rotation
    when the maximum value of 20\%\ polarization at 576 nm is measured, almost the 
    entire visible stellar surface is covered by magnetic fields between $1.4\cdot 10^4$
    and $2.0\cdot 10^4$ T, with only a small tail extending to maximum field 
    strengths of $5.3\cdot 10^4$. This distribution is best reproduced at a rotational
    phase where the axis of the off-centered dipole is nearly perpendicular to the line
    of sight. Using this field geometry 
    we calculated a histogram distribution of the visible surface magnetic field 
    strengths in order to set sharp limits on gravitational birefringence.     
    For each field strength bin of the histogram we calculated the maximum circular
    polarization from radiative transfer calculations by assuming that the field vector 
    always points towards the observer. The total maximum polarization from the whole 
    visible stellar disk without gravitational birefringence is then calculated by 
    adding up the contributions from each field strength bin weighted with its relative 
    frequency. This results in $V_{\lambda, {\rm max}}/I_{\lambda,{\rm tot}} = 26.5$\%. 
    Assuming that the reduction to $V_{\lambda, {\rm obs}}/I_{\lambda,{\rm tot}}= 20$\% 
    is entirely due to gravity induced depolarization -- and not due to the fact that 
    in reality not all field vectors point towards the observer -- we find an upper 
    limit for this effect of $p^2 \lsim (0.9\,{\rm km})^2$. Since there is always a small
    uncertainty in determing the exact mass of a white dwarf, we also calculated an upper 
    limit on $p^2$ assuming a lower mass of $1\,M_{\odot}$. This leads to 
    $p^2 \lsim (1.2\,{\rm km})^2$.    
    An even more extreme assumption would be to take 100\%\ emerging polarization, i.e.
    neglect the dipole model and make no reference to radiative transfer calculations. 
    This leads to $p^2 \lsim (2.125\,{\rm km})^2$.
    
  \section{Discussion and Conclusions}
    We have shown that the Poincar\'e gauge theory exhibits gravitational birefringence
    under the assumption of a specific nonminimal coupling and have given an explicit expression
    for the gravity-induced phase shift between orthogonal polarization states. Using
    spectropolarimetric observations of the massive white dwarf RE J0317-853 we imposed
    strong constraints on the birefringence of spacetime with an upper limit on the
    relevant coupling constant $p^2$ of $(0.9\,{\rm km})^2$ or $p^2 \lsim (2.125\,{\rm km})^2$
    for the most conservative assumptions. Since gravity-induced birefringence violates
    the Einstein equivalence principle, our analysis also provides a test of this foundation
    of general relativity. Tighter limits could be achieved either by observing more massive
    white dwarfs or by circular polarization measurements at significantly shorter wavelength,
    such as in the far ultraviolet (e.g. in the Ly$\alpha$ absorption features). In addition,
    a consistent model for the magnetic field geometry which reproduces the spectropolarimetric
    measurements in the optical would help.

    The properties of the exchange particles of the torsion field within PGT, especially their
    masses, are currently not bound from the theoretical side and, therefore, the relevance
    for astrophysical observations still requires further work \cite{blago,baek86}.

\end{document}